\documentclass[a4paper,twocolumn,twoside,11pt]{article}

\usepackage[T1]{fontenc}
\usepackage{indentfirst}
\usepackage{abstract}
\usepackage[hang,flushmargin]{footmisc}
\usepackage{titlesec}
\titleformat{\section}%
  {\centering\expandafter\MakeUppercase\expandafter}{\thesection.}{0.5em}{}
\titleformat{\subsection}%
  {\centering\slshape}{\thesubsection.}{0.5em}{}
\titleformat{\subsubsection}[runin]%
{\normalfont\normalsize\bfseries}{\thesubsubsection.}{1em}{\addperiod}
\newcommand{\addperiod}[1]{#1.}
\usepackage{caption}

\makeatletter
\newcommand{\nextauth}{%
 \@ifnextchar[{\nextauth@email}{\nextauth@nothing}%
}
\newcommand{\nextauth@nothing}[2]{%
  \phantom{x}#1$^{#2}$\hspace{-3pt}
}
\newcommand{\nextauth@email}[3][]{%
\if\relax\detokenize{#1}\relax
 \phantom{x}#2$^{#3}$
\else
 \phantom{x}#2$^{#3}$\thanks{\emailname: \texttt{#1}}\,\,%
\fi
}
\makeatother

\newcounter{affil}
\newcommand{\nextaffil}[1]{%
  \stepcounter{affil}
  $^{\arabic{affil}}$\textit{#1}\\
}

\newcommand{\affiliations}[1]{
\date{#1}
}

\newcommand{\keywords}[1]{%
\phantom{x}\\
\noindent\textit{{\keywordsname}:} #1}
\newcommand{\doi}[1]{%
\phantom{x}\\
\noindent\textbf{DOI:} #1}

\def\emailname{E-mail}
\def\keywordsname{Keywords}

\def\aap{\rm{Astron.\ Astrophys.}}

\def\apj{\rm{Astrophys.~J.}}

\def\apss{\rm{Astrophys.\ Space Sci.}}

\def\mnras{\rm{MNRAS}}

\def\prd{\rm{Phys.\ Rev.\ D}}

\usepackage{graphicx}
\usepackage{latexsym,amssymb,amsmath,color}
\usepackage{hyperref}
\usepackage{natbib}

\newcommand{\Mdot}{\dot{M}}

\begin{document}

   \author{\parbox{.9\linewidth}{\centering%
     \textbf{%
  \nextauth[markozoviv@mail.ru]{I. D. Markozov}{1},
  \nextauth{A. D. Kaminker}{1},
  \nextauth{A. Y. Potekhin}{1}
     }}}

\title{\expandafter\MakeUppercase\expandafter{\bf{Hydrodynamical simulation 
of the structure of the X-ray pulsar accretion channel: accounting for resonant scattering}}}

\markboth{\small\rm \hfill MARKOZOV et al. \hfill}{}
\markright{\expandafter\MakeUppercase\expandafter{\small\hfill\rm Accretion channel
 with resonant scattering \hfill}}

\affiliations{
\nextaffil{Ioffe Institute, Politekhnicheskaya 26, St.~Petersburg, 194021 Russia}
}

\def\abstractname{\vspace*{-8ex}}

\twocolumn[
   \begin{onecolabstract}    
 \maketitle
   \noindent\normalsize
A self-consistent radiation-hydrodynamics model of an accretion channel
of subcritical X-ray pulsars is constructed.
The influence of the presence of resonance in the scattering cross-section on the accretion process and radiation
transfer is taken into account. It is shown that the efficiency of plasma deceleration by radiation depends on the
magnitude of the magnetic field $B$. For $B=1.7\times 10^{12}$~G,
the spectra and the degree of linear polarization of the radiation 
of the accretion channel are constructed. In the obtained spectra, the shape of the cyclotron line depends
on the direction of the outgoing radiation. The calculated linear
polarization  degree
 of the outgoing radiation is  $30 -40\%$ near the cyclotron resonance, 
 whereas it can be small ($\lesssim  5 - 10\%$) at energies significantly lower than
the resonant one.
\vspace{\baselineskip}
\end{onecolabstract}
]
\saythanks 


\section{Introduction}
\label{introduction}

Accreting X-ray pulsars have magnetic fields $10^{11}-10^{13}$~G
\citep[see, e.g., the review by][]{MushtukovTsygankov22}.  
When the plasma moving to such a pulsar reaches the magnetosphere, it
freezes 
into the magnetic field and moves along it to the magnetic poles of the neutron star. The plasma
in the accretion channel near the poles moves with moderately relativistic velocities almost
perpendicular to the surface \citep{Davidson73}. The kinetic energy of the plasma is converted into
radiation, whose pressure can be so strong that it affects the dynamics of the infalling matter. The
higher the accretion rate, the higher the luminosity. If a certain threshold accretion rate is exceeded,
the radiation can completely halt the matter. In this case, a radiation-dominated shock wave is expected
to appear in the channel \citep{BaskoSunyaev76}, behind the front of which a region of slow
sedimentation of matter is formed. Hereafter, X-ray pulsars of this type will be called
\emph{supercritical}, while those with an accretion rate below the threshold will be called
\emph{subcritical}.

A large amount of the latest observational data on X-ray pulsars leads to the importance and relevance
of theoretical modeling of the structure and radiation of these objects. Meanwhile, currently there is
no sufficiently complete theoretical model capable of describing all the variety of physical processes
in a wide range of parameters of the X-ray pulsars. Since the radiation generated by accretion actively
affects the accretion process itself, theoretical modeling of the structure of the accretion channel and
the characteristics of its radiation should be carried out in a self-consistent manner. An example of
such calculations in the stationary case has been presented  by \citet{West_17a,West_17b}. Nonstationary
modeling without calculation of spectra was first performed by \citet{KleinArons89}. They have
demonstrated the formation of a radiation-dominated shock wave in the accretion channel of a
supercritical pulsar. A one-dimensional calculation of the process of establishing a stationary flow in
the accretion columns of supercritical pulsars was carried out by \citet{AbolmasovLipunova23}, who have
taken into account the possibility of the column leakage at highly supercritical accretion rates and
found the limits of applicability of the analytical solution of \citet{BaskoSunyaev76}. The simulations
of X-ray pulsar radiation with an accurate account of the magnetic field effect on radiative transfer
in the plasma were carried out separately from the solution of the equations of hydrodynamics. The most
detailed calculations to obtain the spectra of X-ray pulsars with cyclotron features were performed
using the Monte Carlo method by \citet{Schwarm_17}.

In this paper, we consider subcritical X-ray pulsars with an accretion channel completely filled with
plasma. For such systems, we present the results of self-consistent radiation-hydrodynamical modeling of
the channel structure and the its outgoing radiation. An important difference between our work and the
previous ones is the joint calculation of radiative transfer and accretion hydrodynamics, taking into
account birefringence and resonance scattering in a magnetic field. In addition, the method we use does
not employ the diffusion approximation, which allows us to consider subcritical pulsars with low density
of matter in accretion channels.

\section{Magnetic field effects}
\label{magnetic}

Quantization of electron motion across magnetic field lines \citep[e.g.,][]{SokolovTernov} can be
important in the accretion channels of the X-ray pulsars. Electrons occupy energy levels (Landau levels),
each of which corresponds to a certain value of the transverse momentum. In this case, the total energy
of the electron with a momentum along the field $p_z$ at the  $n$th Landau level ($n=0,1,2,\ldots$)
equals
\begin{equation}\label{markozov:eq1}
E_n=\sqrt{m^2c^4+c^2p_z^2+2nmc^2E_\mathrm{cyc}},
\end{equation}
where $E_\mathrm{cyc}={\hbar eB}/{(mc)}$ is the cyclotron energy, $e$ is the electron charge, $m$ is its
mass, $B$ is the magnetic field strength, $\hbar$ is the reduced Planck constant, and $c$ is the speed
of light.  We assume that all electrons 
occupy
the ground Landau level, and consider transitions only
from the ground level to the ground level, which is justified by the short lifetime of electrons at
excited Landau levels compared to the characteristic free path time of an electron in the pulsar
magnetosphere \citep[e.g.,][]{Meszaros92}.

The magnetized plasma is a birefringent medium: the radiation splits in it into two waves, extraordinary
and ordinary ones \citep[e.g.,][]{Ginzburg,GnedinPavlov74}, which are often called the X-mode and the
O-mode. In general, they have elliptical polarization; the major semiaxis of the ellipse that is the
locus of the endpoints of the electric vector of the O-mode lies in the plane formed by the magnetic
field vector and the photon wave vector, while the major semiaxis of the ellipse corresponding to the
X-mode is perpendicular to this plane.

The ellipticity of normal modes depends on the photon energy and propagation direction. In this paper,
it is calculated without allowance for vacuum polarization and temperature effects in the plasma
\citep[see, for example, fig.~1 and eq.~(3) in][]{Mushtukov_22}. 
In this case, for the applicability of the
normal mode approximation it is sufficient that $\min(\omega,\omega_\mathrm{cyc})\gg \nu_e$
\citep{GnedinPavlov74}, where $\omega$ is the photon frequency, $\omega_\mathrm{cyc}$ is the electron
cyclotron frequency, and $\nu_e$ is the collision frequency. 
The frequencies of radiative and non-radiative
collisions of electrons with protons in quantizing magnetic fields are given in the article by
\citet{PotekhinLai07}. Using them, it is easy to verify that the condition for the applicability of the
normal mode approximation in the accretion channel under consideration is fulfilled with a large margin.
For example, neglecting Coulomb logarithms for order-of-magnitude estimates and setting
$B=2\times10^{12}$~G, $\rho=10^{-5}$~g/cm$^3$, and $E=T=1$~keV, we obtain $\nu_e/\omega \sim 10^{-10}$.

The cross sections of radiation processes in a strong magnetic field depend on polarization. In
addition, they have resonances, which cause the appearance of cyclotron lines in the X-ray pulsar
spectra. We will consider only the processes of Compton scattering of photons by electrons in a strong
magnetic field. In this case, the laws of conservation of energy and longitudinal momentum are
fulfilled, while the transverse momentum is not conserved. For photons of the normal modes, we use
approximate expressions for scattering cross sections obtained by \citet{Herold79}, which contain only
the main resonance at the cyclotron energy $E_\mathrm{cyc}$. To calculate birefringence, we use the cold
plasma approximation \citep[see][]{Ginzburg,GnedinSunyaev74}, which does not take into account the
electron thermal motion effect on  the  dielectric tensor. We also neglect the vacuum polarization
\citep{PavlovGnedin84}. Full expressions for scattering cross sections in the representation of elliptic
modes can be found in the paper by \citet{Mushtukov_22}. When averaging cross sections over an ensemble
of electrons, we used the relativistic Maxwell distribution with a temperature of $T=5$~keV. Such an
approach, in which $T$ is not calculated in a self-consistent manner, gives only qualitative results for
radiation energies in the resonance region.

\section{Statement of the problem}
\label{Equations}

We describe plasma motion in the accretion channel by non-relativistic equations of radiation
hydrodynamics \citep[see][]{Castor},
	\begin{equation}\label{markozov:eq2}
      \left\{
      \begin{array}{l}
         \displaystyle
          \frac{\partial\rho}{\partial{t}}+\nabla\cdot(\rho\mathbf{v})=0, \\[2ex]
          \displaystyle
          \frac{\partial\rho\mathbf{v}}{\partial{t}}+\nabla\cdot(\rho\mathbf{v}\otimes\mathbf{v})+\nabla{p}=\mathbf{F}_{g}+\mathbf{F}_{r}, \\[2ex]
          \displaystyle
          \frac{\partial}{\partial{t}}(\rho{\epsilon}+\frac{1}{2}\rho{v^2})+\nabla\cdot(\rho\mathbf{v}h+\frac{1}{2}\rho\mathbf{v}v^2)=Q_g+Q_r.
      \end{array}
      \right.
    \end{equation}
 Here, $\rho$ is the mass density, $\mathbf{v}$ is the plasma velocity,  $p$ is the pressure,
$\rho\epsilon$ is the internal energy density, and $\rho h$ is the enthalpy density of matter;
$\mathbf{F}_{g}=\rho\mathbf{g}$ is the gravitational force density, $Q_g=\rho\mathbf{v}\cdot\mathbf{g}$
is its power, and $\mathbf{g}$ is the gravitational acceleration. We neglect the General Relativity
effects, hence $g={GM}/{r^2}$, where $M$ is the mass of the star, $r$ is the distance to its center, and
$G$ is the Newtonian constant of gravitation. The term $Q_r=-\int dE\int
d\Omega\left(\varepsilon_E-\alpha_EI_E\right)$ is the power density of energy exchange between the
plasma and radiation, $\mathbf{F}_{r}=-\frac{1}{c}\int dE\int
d\Omega\mathbf{\Omega}\left(\varepsilon_E-\alpha_EI_E\right)$ characterizes their momentum exchange,
where $\alpha_E$ is the absorption coefficient of photons with energy $E$,
$\varepsilon_E$ is their emission coefficient in the medium,  $\mathbf{\Omega}$ is a unit vector of photon propagation
direction, $d\Omega$ is a solid angle element, and $I_E$ is the specific intensity\footnote{
Quantities $I_E$ and $\varepsilon_E$ are normalized to the unit photon energy interval:
$I_E=I_\nu/(2\pi\hbar)$, $\varepsilon_E=\varepsilon_\nu/(2\pi\hbar)$, where $I_\nu$ and
$\varepsilon_\nu$ are the specific intensity and emission coefficient, normalized to the unit frequency
interval \citep[e.g.,][]{Sobolev}.}. To calculate $I_E$, $Q_r$, and $\mathbf{F}_r$, one needs to solve
the equation of radiative transfer in the medium,
\begin{multline}\label{markozov:eq3}
    	\mathbf{\Omega}\cdot\mathbf{\nabla}I^{m}_E=\varepsilon_E^m-\alpha_E^mI_E^m
        \\
        =\sum\limits_{q=1}^{2}\int\limits_0^\infty{dE'\int\limits_{4\pi}d\Omega'}
        \big[R_{mq}(E,\mathbf{\Omega}|E',\mathbf{\Omega}')I_{E'}^{q}(\mathbf{\Omega}')\\
	-R_{qm}(E',\mathbf{\Omega}'|E,\mathbf{\Omega})I_{E}^{m}(\mathbf{\Omega})\big] .
 \end{multline}
 Here, subscripts $m$ and $q$ denote photon polarization ($m,q = 1$ and 2 for the X- and O-mode,
respectively), $R_{mq}(E,\mathbf{\Omega}|E',\mathbf{\Omega}')$ is the scattering coefficient for
photons with energy $E'$, which move in the direction $\mathbf{\Omega}'$ and have polarization $q$,
into the state with energy $E$, direction $\mathbf{\Omega}$, and polarization $m$. We neglect true
absorption and emission, as well as stimulated processes, and consider only scattering in which the
photon is preserved. The total specific intensity and the total emission coefficient are given by the
sum of the polarizations: $I_E=\sum_{m=1}^2I_E^m$, $\varepsilon_E=\sum_{m=1}^2\varepsilon_E^m$. 

    One needs to set boundary conditions to the system of radiation hydrodynamics equations
(\ref{markozov:eq2}). We assume that the matter at the upper boundary of the channel is in the free-fall
state and at a certain fixed accretion rate $\dot{M}$ its velocity is $v=\sqrt{\frac{2GM}{R+H}}$, where
$R$ is the neutron-star radius and $H$ is the height of the channel. We neglect the gas pressure at the
upper boundary. It does not significantly affect the resulting solutions, since the flow dynamics weakly
depends on the gas pressure.

 The problem of choosing boundary conditions at the neutron-star surface is less trivial. This problem
was discussed, for example, in the works  by \citet{BaskoSunyaev76} and \citet{Kirk84}. Here we choose
the simplest type of boundary condition: $\dot{E}_r=\dot{E}_k+\dot{E}_\mathrm{back}$, where
$\dot{E}_k$ is the kinetic energy of matter flowing into the lower boundary per unit time,
$\dot{E}_\mathrm{back}$ is the power of radiation scattered from the accretion channel to the lower
boundary, and $\dot{E}_r$ is the power of  radiation emitted from the lower boundary and having a
Planck spectrum.  It is also assumed  that all matter is in the free-fall state at the initial  moment
of time.

 \section{Method of solution}
 \label{sol_method}

The joint system of equations of radiation hydrodynamics and radiation transfer was solved using a
splitting scheme. The time step was divided into two substeps. At the first substep, the hydrodynamic
equations were solved without radiation.  This was done using the VH-1 library (``Virginia Hydrodynamics
1'', \href{http://wonka.physics.ncsu.edu/pub/VH-1}{http://wonka.physics.ncsu.edu/pub/VH-1}). It
implements a piecewise parabolic method of the third order of accuracy with a transition to a Lagrangian
grid \citep[PPMLR, developed by][]{ColellaWoodward84}, which belongs to the class of Godunov methods. 
At the second  substep the  Monte Carlo method was used to calculate  the radiation transfer. The
distributions obtained  at the first substep were used as the values of density and velocity  in the
accretion channel at the second substep. During the elementary scattering process, the change  in plasma
energy and momentum at a given point was calculated.  Further, according to the known changes in these
values for the entire  radiation substep, the final values of plasma pressure and velocity were
recalculated.

\begin{figure}
    \centering
    \includegraphics[width=\columnwidth]{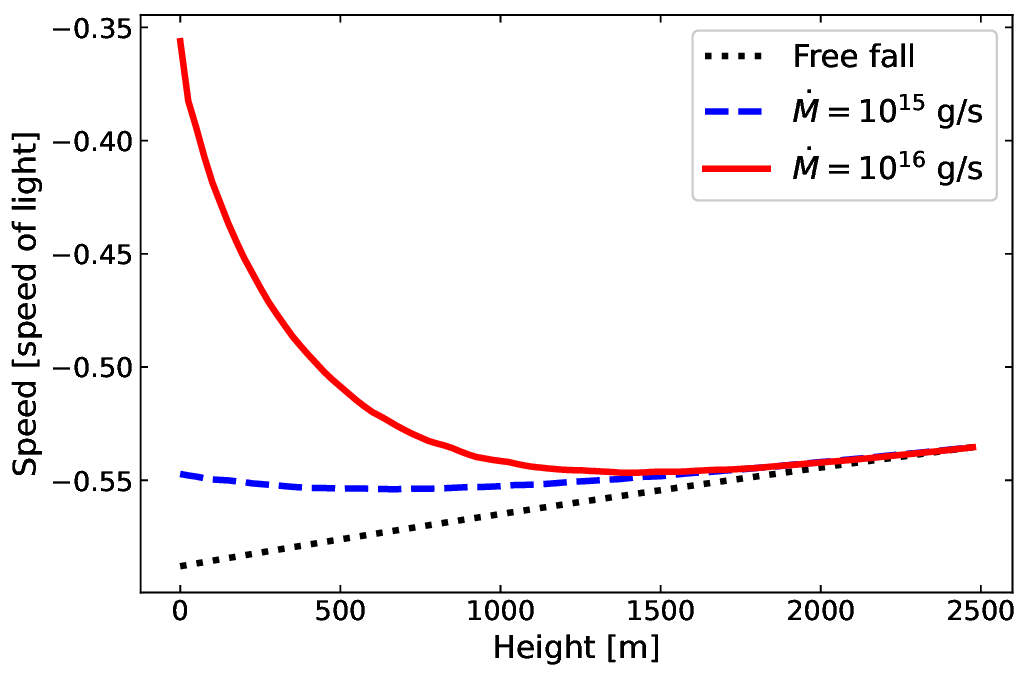}
    \caption{Accreting matter velocity profiles in units of the speed of light as functions of the
height above the neutron-star surface. The black dotted line corresponds to free fall, the blue dashed
line corresponds to the accretion rate $\dot{M} = 10^{15}$~g/s, and the red solid line corresponds to
$\dot{M} = 10^{16}$~g/s. The parameters are $M = 1.4M_\odot$, $R = 12$~km, $R_c=1$~km, $H=2.5$~km; the
Thomson scattering cross section is used. All values are averaged along the radial coordinate in the
accretion channel  and correspond to steady-state currents.}
    \label{Velocity_1}
    \end{figure}

To account for Compton scattering, we generated tables of the cumulative distribution function
$f_{mq}(E_i,\theta_i,\theta_f)$ for the probability that a photon with 
energy $E_i$ and polarization $q$, initially moving at angle $\theta_i$, to the magnetic field will get polarization $m$  and angle
$\theta_f$ to the magnetic field after scattering.  The tabular values $f$ were interpolated for
arbitrary parameter values in the rest  frame of scattering electrons. The accreting matter moves at a
certain speed, so first a transition is made from the reference frame of the neutron star to the
(moving) rest  frame of the plasma. Photon energy $E_i^p$ and angle $\theta_i^p$  are calculated in the
latter reference frame according to the formulae of the Lorentz transformation. The value of the angle
after scattering in the plasma reference frame is  obtained as
$\theta_f^p=f^{-1}(E_i^p,\theta_i^p,\eta)$, where $\eta$  is a generated random value with uniform
distribution and $f^{-1}$ is  the inverse function. The angle in the reference frame of a neutron star
is obtained using the inverse Lorentz transformations.  The photon energy after scattering at a known
angle is calculated  according to the laws of energy and momentum conservation (where, in the considered
case of $n=0$, it is sufficient to take into account only the longitudinal momentum of the electron,
$p_z$).

We neglect the curvature of the magnetic field and assume that the accretion 
channel has the shape of a cylinder.
This approximation is justified when
the heights at which the radiation pressure affects the plasma dynamics
are small compared to the radius of a neutron star. The channel
was divided into slices of equal height, and in the transverse coordinate
(the radius measured from the axis of the cylinder) 
it was divided into rings of equal areas. We assume that the plasma
is completely frozen in a magnetic field.
Then there is no macroscopic motion of matter across the field lines,
and the two-dimensional hydrodynamic problem turns into a series of one-dimensional ones: 
a separate calculation is performed for each ring. 
On the contrary, the radiation transfer was calculated in the completely three-dimensional form.

 \begin{figure}
    \centering
    \includegraphics[width=.93\columnwidth]{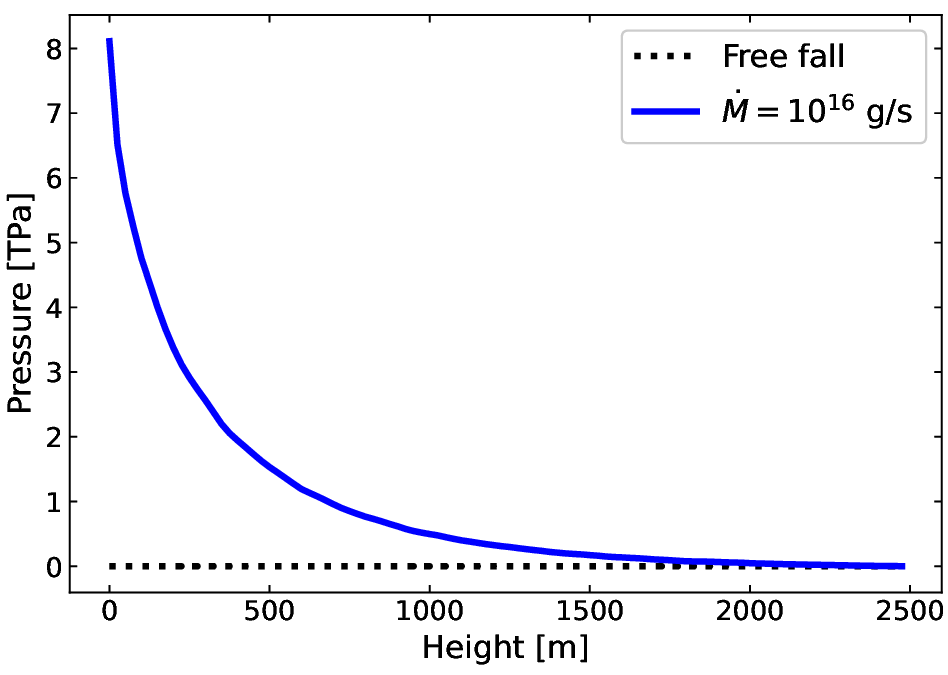}
    \caption{Pressure profiles in the accretion channel (in units of $10^{12}$~Pa),  averaged along the
radial coordinate (of the cylinder), as functions of height above the neutron-star surface. The black
dotted line corresponds to free fall and the solid blue one to the steady flow with accretion rate
$\dot{M} = 10^{16}$~g/s.}
    \label{Pressure}
    \end{figure}

   \begin{figure}
    \centering
    \includegraphics[width=\columnwidth]{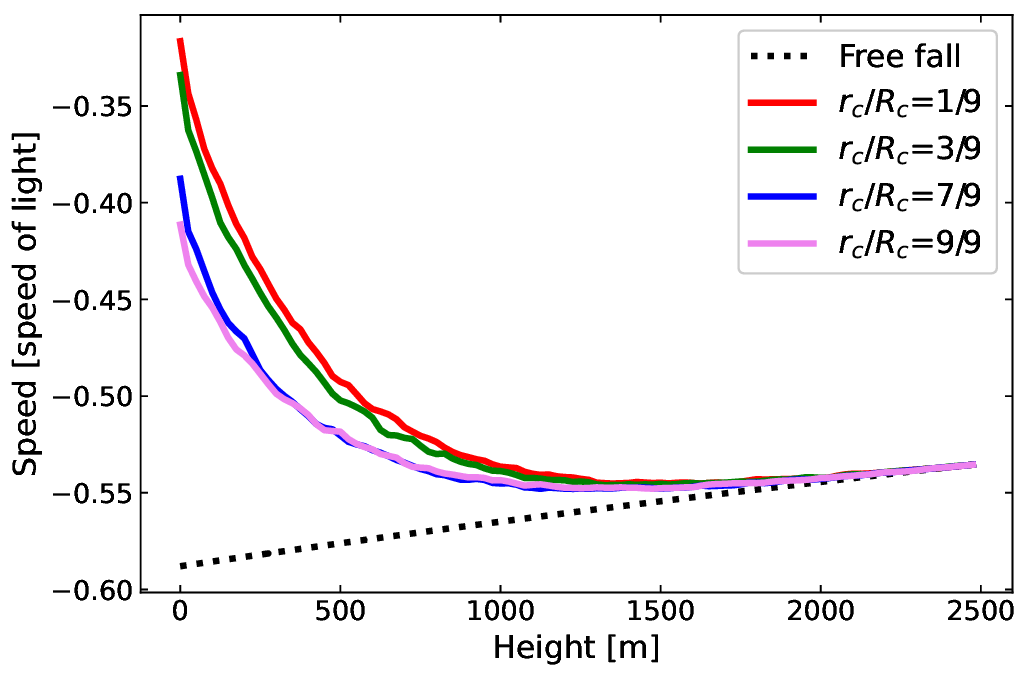}
    \caption{Velocity profiles of matter in units of the speed of light as functions of height above the
neutron-star surface at different distances $r_c$ from the axis of the accretion channel. The red
(upper) curve -- $r_c/R_c=1/9$ (central parts of the channel),  the green curve -- $r_c/R_c=3/9$, the
blue curve -- $r_c/R_c=7/9$,  the purple curve --  $r_c/R_c=9/9$ (the edge of the channel).   Accretion
rate is $\dot{M} = 10^{16}$~g/s,  and the Thomson scattering cross section is used.  The distributions
correspond to steady-state currents.} 
    \label{Velocity_Tr}
    \end{figure}

   \section{Numerical simulation results}

The main parameters of the model are the mass of the neutron star $M$, 
its radius $R$, the accretion rate $\dot{M}$, the radius of the 
accretion channel $R_c$, its height $H$, and the cyclotron energy 
$E_\mathrm{cyc}$. We considered a neutron star with mass $M=1.4M_\odot$ 
and radius $R=12$~km and an accretion channel with radius $R_c=1$~km 
and height $H=2.5$~km.

    \subsection{Hydrodynamics}
    \label{results:hydro}
     
In our calculations, the evolution of hydrodynamic characteristics was monitored  until the steady flow
of plasma in the channel was established. Figures \ref{Velocity_1} and \ref{Pressure} show dependences
of velocity and pressure  on height above the surface of a neutron star for steady-state flows.  The
negative sign of the velocity means that the motion is towards the surface. The values are averaged
along the radial coordinate in the cylinder, which corresponds to one-dimensional modeling.  It can be
seen from the figures that the matter slows down near the surface,  the deceleration being the stronger,
the greater the accretion rate.  It is caused by the radiative pressure on the accreted matter.  The
radiation is generated as a result of the impact of the incident plasma on the neutron-star surface. The
deceleration of matter occurs on the scale of $\sim1$ -- 2~km. This is much smaller than the radius of
the star, which justifies the cylindrical approximation for the accretion channel.

  \begin{figure}
    \centering
    \includegraphics[width=\columnwidth]{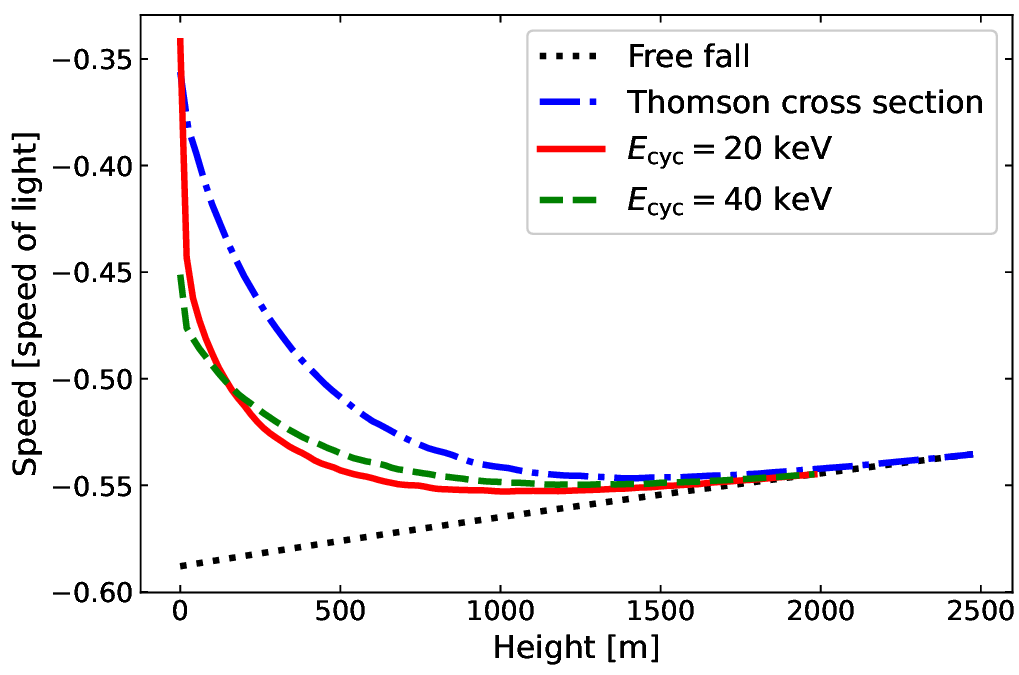}
    \caption{Velocity profiles in units of the speed of light as functions of
height above the neutron-star surface 
for the case of a non-magnetic Thomson cross section (blue dot-dashed line), 
magnetic cross section with $E_\mathrm{cyc}=20$~keV (red solid line) and
 $E_\mathrm{cyc}=40$~keV (green dashed line). 
 The black dotted curve corresponds to the state of free fall. 
 All values are averaged along the radial coordinate in the accretion channel 
 and correspond to steady-state currents.}
    \label{Velocity_2}
    \end{figure}

  \begin{figure*}[t]
    \centering
    \includegraphics[width=\columnwidth]{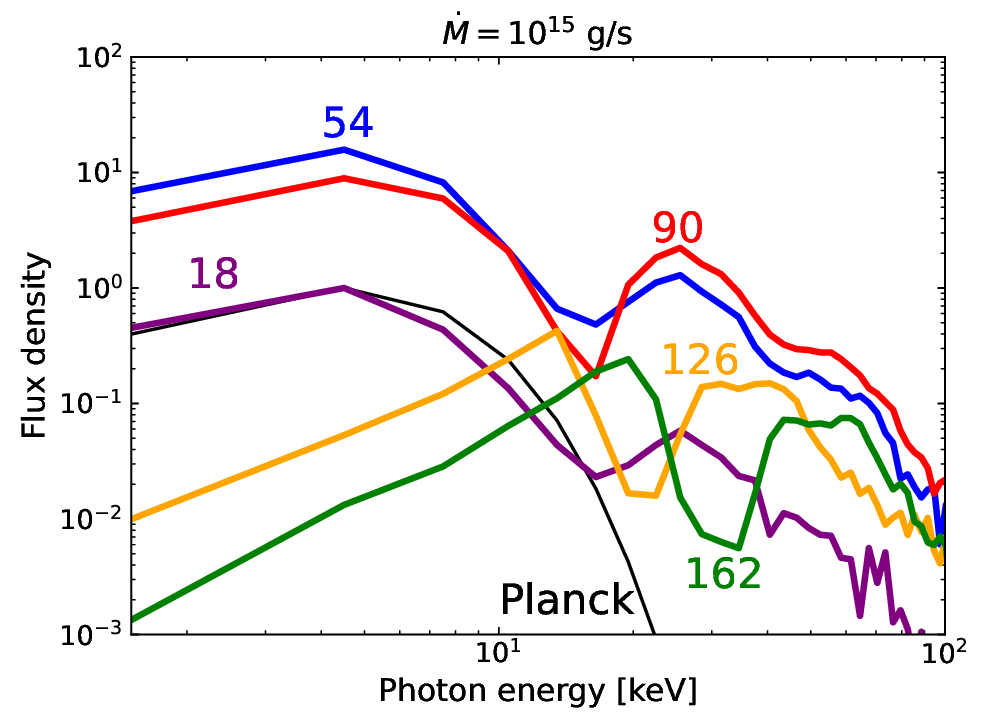}\,\,
     \includegraphics[width=\columnwidth]{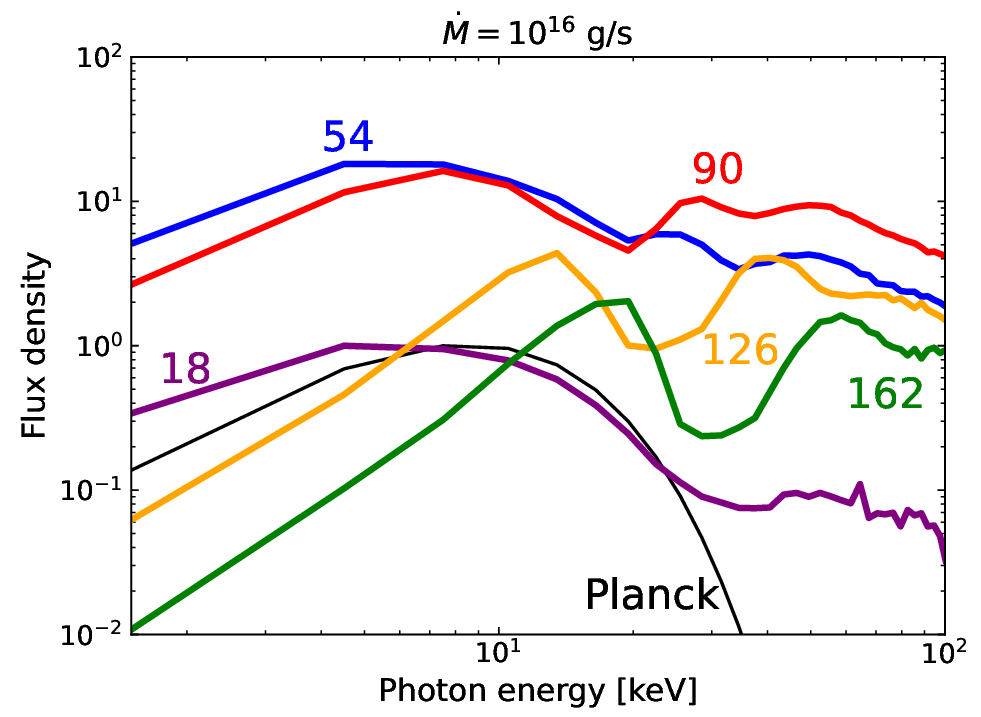}
    \caption{
  The average density of the photon energy flux in the range of 
  directions of $0.02\pi$ around the central values $\theta$, indicated 
  by the numbers near the curves (in degrees), as a function of the 
  energy of photons in a magnetic field with $E_\mathrm{cyc} = 20$~keV. 
  The normalization is to the maximum of the curve with 
  $\theta=18^\circ$. Left panel: accretion rate $\dot{M}=10^{15}$~g/s, 
  right panel: $\dot{M}=10^{16}$~g/s. Integration over the azimuthal 
  angle has been performed.}
    \label{Spectrum}
    \end{figure*}

Velocity profiles in the channel for different distances from the axis of the cylinder are shown in
Fig.~\ref{Velocity_Tr}. One can see that the plasma located at the center of the channel experiences the
strongest deceleration, and the deceleration monotonously weakens towards the edges. However, this
effect is not as pronounced as in the case of supercritical accretion \citep[see,
e.g.,][]{Mushtukov_15,Gornostaev21}.

In  Fig.~\ref{Velocity_2} we compare velocity profiles in the accretion  channel obtained for the
Thomson scattering cross section and the cross section in a magnetic field accounting for the cyclotron
resonance. A feature of scattering in a strong magnetic field is the presence of a sharper velocity
gradient in the channel areas, close to the surface of a neutron star. That is, a stronger deceleration
due to resonant processes occurs at lower altitudes  than in the case of Thomson scattering. Indeed, in
the resonance region, the free path length a relatively small part of photons with energies $E \sim
E_\mathrm{cyc}$  is greatly reduced, leading to a decrease of the effective height of the deceleration.
Nevertheless, at the cyclotron energy $E_\mathrm{cyc}=20$~keV ($B=1.7\times 10^{12}$~G)  and the
temperature of the boundary surface emitting the Planck spectrum $T=3$~keV, plasma velocity $v$ at the
very neutron-star surface turns out to be approximately equal to $-0.35\,c$, as in the case of Thomson
scattering. However, for still stronger magnetic fields, a sharp deceleration of matter at the very
surface of the star is noticeably smaller than in the case of $E_\mathrm{cyc}=20$~keV. For instance,
Fig.~\ref{Velocity_2} shows the accretion velocity profile at $E_\mathrm{cyc}=40$~keV ($B=3.4\times
10^{12}$~G)  and the lower boundary temperature $T=2.8$~keV,  when the velocity at the surface of the
star is $v\approx - 0.45\,c$.  In this case, the energies of the majority of  photons $E \sim T$ are
shifted more strongly to the region of $E\ll E_\mathrm{cyc}$, where scattering cross sections are
suppressed by a small factor $(E/E_\mathrm{cyc})^2$, while the number of resonant photons (that provide
the braking) decreases significantly, which leads to a relative increase of the accreting plasma
velocity.

 Note that the comparison of the structure of accretion channels of subcritical pulsars with the
scattering in strong magnetic fields and with the Thomson scattering  agrees with the results of a
similar comparison in the work of \citet{Sheng_23} for a supercritical accretion regime.

    \subsection{Radiation}
    \label{Radiation}

    \begin{figure*}
    \centering
    \includegraphics[width=\columnwidth]{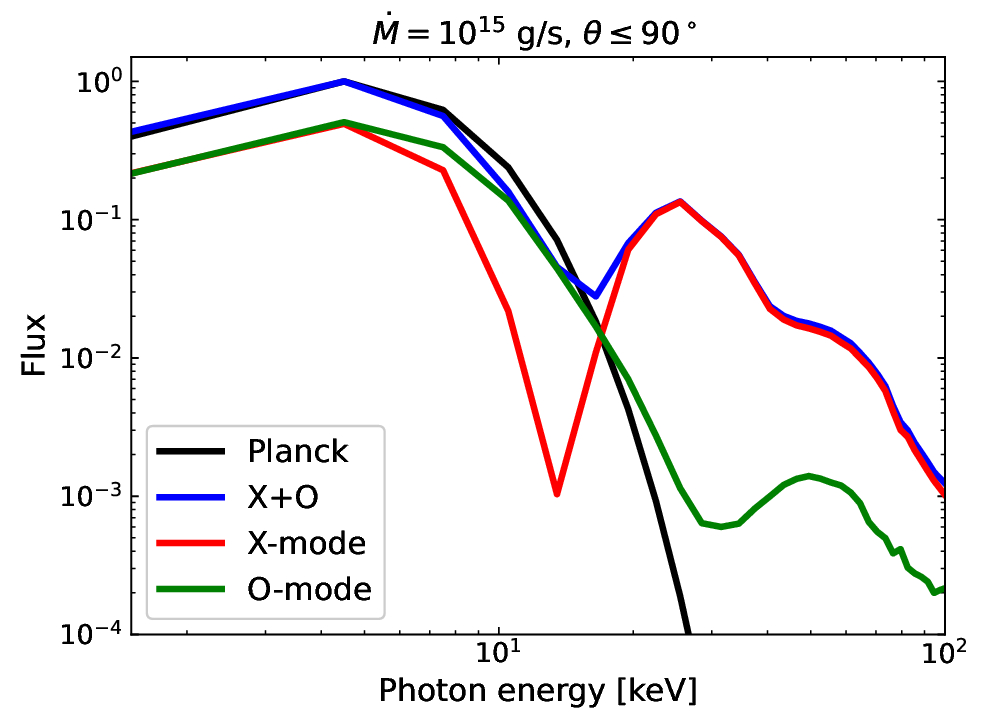}
     \includegraphics[width=\columnwidth]{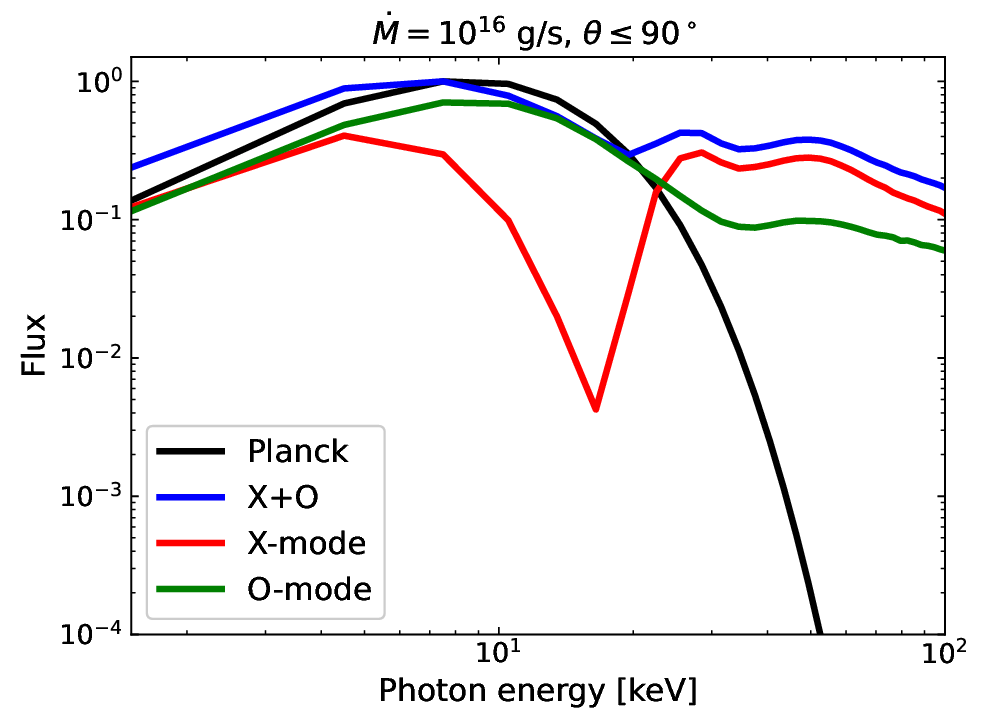}
    \caption{Spectral fluxes coming out of the accretion channel, 
    as functions of the photon energy
for radiation in the X-mode (red curve), O-mode (green curve), and the sum of X+O modes (blue curve).
The black curve corresponds to the Planck spectrum.
The radiation is integrated over angles $\theta\leq90^\circ$, 
  normalization is to the maximum of the sum of the two modes.
    }
    \label{Spectra_first_half}
    \end{figure*}
    
 Along with the distribution of hydrodynamic quantities, we also simulate the characteristics of
radiation coming out of the accretion channel. Fig.~\ref{Spectrum} presents the spectral (over the
photon energy) distributions of flux densities of the photon energy around different directions
$\theta$. Here the angle $\theta$ is measured from the outer normal to the neutron-star surface, that
is, the value $\theta=90^\circ$ corresponds to the direction perpendicular to the channel walls. The
cyclotron resonance corresponds to the energy $E_\mathrm{cyc}=20$~keV.

    The spectra reveal cyclotron absorption lines, which are most pronounced for the angles
     $\theta>90^\circ$. Photons coming out in such directions have experienced at least 
     one scattering and are directed mainly towards the surface of the neutron star.
Due to the relativistic Doppler effect
the position of the cyclotron line depends on the angle at which the radiation exits.
    
    Fig.~\ref{Spectra_first_half} presents spectral fluxes in the range of angles
    $\theta\leq90^\circ$. 
    In this case, the radiation propagates directly towards the observer 
    and does not cross the surface of the neutron star. 
    The graphs show that the O-mode dominates at the resonance,
but the X-mode starts to dominate with increasing the photon energy to $E>E_\mathrm{cyc}$.
    
    With a known radiation intensity in the two modes, one can calculate the degrees of linear ($P_L$)
and circular ($P_C$) polarization. In the cold plasma approximation, they have the form
\citep{KaminkerPS82}
    \begin{equation}\label{markozov:P_L}
    P_L=\frac{I_O-I_X}{I_O+I_X}\frac{|q|}{\sqrt{1+q^2}},
    \quad
    P_C=\frac{I_X-I_O}{I_O+I_X}\frac{\mathrm{sign}(q)}{\sqrt{1+q^2}},
    \end{equation}
where $I_X$ is the specific intensity in the X-mode, $I_O$ in the O-mode, 
$q=\frac{E_\mathrm{cyc}}{E}\frac{\sin^2{\theta}}{2\cos{\theta}}$,  $E$ is the photon energy, and
$\theta$ is the angle between the direction of photon and the magnetic field.  

\begin{figure*}
    \centering
    \includegraphics[width=\columnwidth]{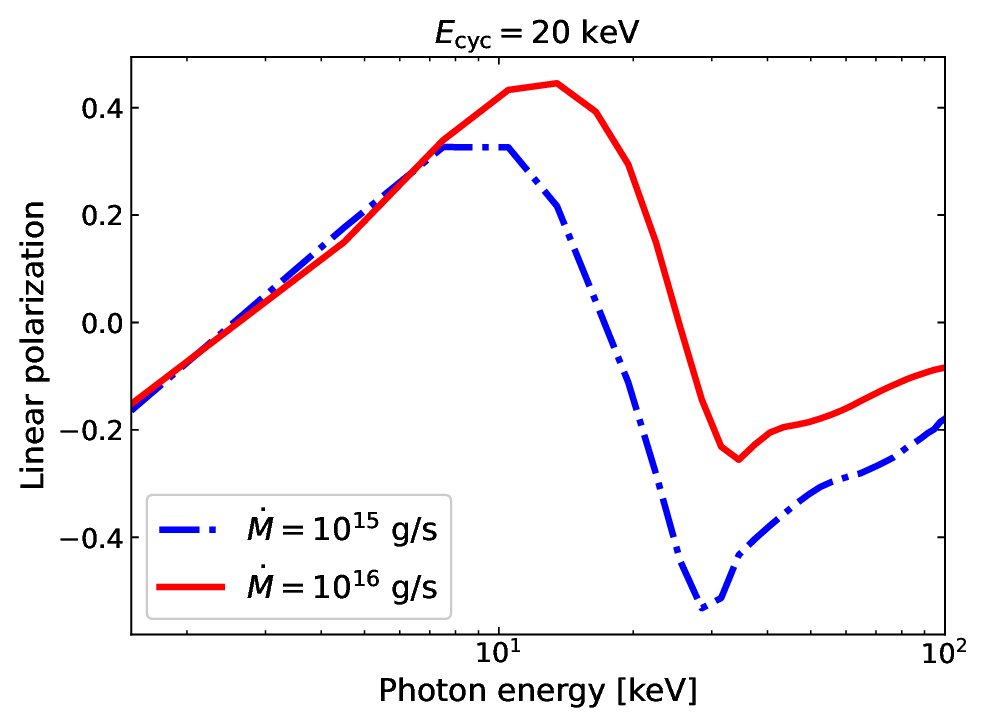}
     \includegraphics[width=\columnwidth]{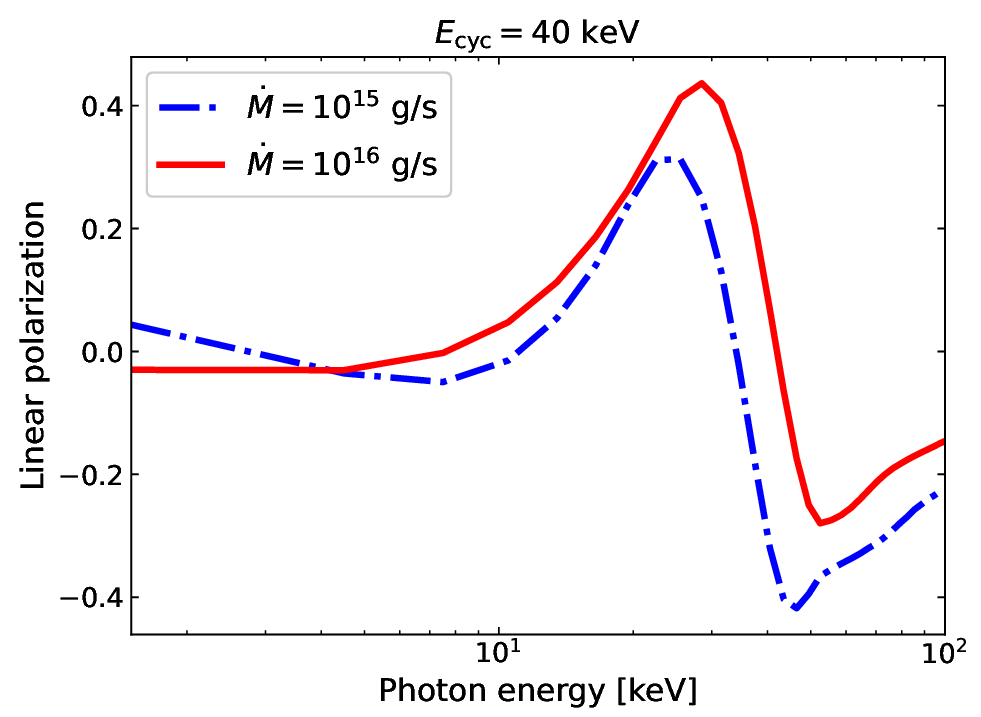}
    \caption{The degree of linear polarization as a function of photon energy for radiation propagating
in the range of angles  $0 \leq \theta \leq 90^\circ$. Left: $E_\mathrm{cyc}=20$~keV,  right:
$E_\mathrm{cyc}=40$~keV. The blue dot-dashed line corresponds to the accretion rate $\Mdot=10^{15}$~g/s,
and the red solid line corresponds to $\Mdot=10^{16}$~g/s.
    }
    \label{Polarization}
    \end{figure*}

    The results of calculation of the degree of linear polarization for cyclotron energies of 20 and
40~keV are shown in Fig.~\ref{Polarization}. In the resonance, the radiation is strongly polarized,
while the degree of polarization can be small ($\lesssim 5 - 10\%$) at lower energies.  In the region of
energies above the resonance, the degree of polarization depends on the accretion rate.    It follows
from Fig.~\ref{Polarization} that the degree of polarization of radiation in a fixed interval of
relatively low energies $E_1 \leq E \leq E_2   <  E_\mathrm{cyc}$ depends on the cyclotron resonance
energy: the larger $E_\mathrm{cyc}$, the smaller the degree of polarization.  Since only a small
fraction of the radiation is subject to scattering at low energies, the low degree of polarization
before the resonance is a consequence of the assumption that the radiation from the lower boundary is
unpolarized. However, this assumption is only the first crude approximation, and the calculation of
polarization with more reliable models of boundary radiation is the subject of further research.

       \section{Conclusion}
    \label{Conclusion}

    A code has been created for self-consistent calculation of the radiation hydrodynamics of matter
flowing along magnetic field lines in the accretion  channel of a subcritical X-ray pulsar and radiation
going out of the channel,  taking into account multiple scattering in a strong magnetic field. The
structure of the plasma flow is modeled taking into account  the resonant scattering of photons on
electrons,  which depends on the state of polarization of the photons. It is shown that characteristic
heights of deceleration of the accretion flow above the neutron-star surface become smaller if one takes
into account the influence of the magnetic field  on the scattering process. We note that the total
deceleration in a sufficiently strong magnetic field can be smaller than in the case of Thomson
scattering.

 The characteristics of radiation going from the accretion channel are calculated.  Spectra of this
radiation reveal cyclotron features, whose shape and position depend on the photon propagation
direction. The strongest cyclotron lines occur in the radiation that propagates towards the neutron-star
surface. Therefore, when constructing a complete model of X-ray pulsar radiation,  it is necessary to
take into account the reflection of the channel radiation  by the star's atmosphere
\citep{Poutanen_13,KylafisTL21}.  A detailed calculation of the radiative transfer for the two modes
allows one to obtain the polarization of X-ray radiation. As a result of the simulation,  it is found
that radiation is strongly polarized at energies close to the resonance: the linear polarization degree
is 30--40\%.  At low energies, polarization degree can be small ($\lesssim 5 - 10\%$), however this is 
a consequence of the chosen boundary conditions, rather than features of radiative transfer in the
accretion channel.

At energies above the resonance, the polarization degree significantly depends on the accretion rate. If
this result will be confirmed in more detailed calculations, then the degree of polarization at
energies  $E>E_\mathrm{cyc}$ can be used as an additional parameter  for determination of the accretion
rate of the X-ray pulsars.

In this paper we did not take into account a number of factors that may have a significant impact on the
obtained results. Despite the characteristic velocities of matter can reach half the speed of light, the
approximation of non-relativistic hydrodynamics was used. In addition, the bremsstrahlung processes of
absorption and emission, the influence of the magnetic field on the spectrum and polarization of
radiation  coming from the surface of the star, and the effects of vacuum polarization were not taken
into account. We are planning to include all these effects sequentially in future calculations.

The work of I.D.M.{} was supported by a grant of the 
Theoretical Physics and Mathematics Advancement Foundation ``BASIS''.


\end{document}